\documentclass{article}
\usepackage{spconf,amssymb,amsmath,graphicx}
\usepackage{epsfig}
\usepackage{xspace}
\usepackage{srcltx}
\usepackage{caption}
\usepackage{subcaption}
\usepackage[export]{adjustbox}
\usepackage{multirow,makecell,hhline}
\usepackage{booktabs}
\usepackage{textcomp}
\usepackage{cite}
\usepackage{hyperref}

\usepackage{color}
\usepackage[dvipsnames]{xcolor}

\newcommand{\CMT}[1]{{}}

\newcommand{\tbh}[1]{\textbf{#1}}

\def\eos{\texttt{</s>} }

\title{Towards Fast and Accurate Streaming End-to-End ASR}
\name{Bo Li, Shuo-yiin Chang, Tara N. Sainath, Ruoming Pang, Yanzhang He, Trevor Strohman, Yonghui Wu}
\address{Google LLC, USA \\
\fontsize{9}{9}\selectfont\ttfamily\upshape
\{boboli,shuoyiin,tsainath,rpang,yanzhanghe,strohman,yonghui\}@google.com}
\begin{document}
\ninept
\maketitle
\begin{abstract}
End-to-end (E2E) models fold the acoustic, pronunciation and language models of a conventional speech recognition model into one neural network with a much smaller number of parameters than a conventional ASR system, thus making it suitable for on-device applications. For example, recurrent neural network transducer (RNN-T) as a streaming E2E model has shown promising potential for on-device ASR \cite{Ryan19}. For such applications, quality and latency are two critical factors. We propose to reduce E2E model's latency by extending the RNN-T endpointer (RNN-T EP) model \cite{Shuoyiin19} with additional early and late penalties. By further applying the minimum word error rate (MWER) training technique \cite{rohit2018mwer}, we achieved 8.0\% relative word error rate (WER) reduction and 130ms 90-percentile latency reduction over \cite{Shuoyiin19} on a Voice Search test set. We also experimented with a second-pass Listen, Attend and Spell (LAS) rescorer \cite{sainath2019two}. Although it did not directly improve the first pass latency, the large WER reduction provides extra room to trade WER for latency. RNN-T EP+LAS, together with MWER training brings in 18.7\% relative WER reduction and 160ms 90-percentile latency reductions compared to the original proposed RNN-T EP \cite{Shuoyiin19} model.  
\end{abstract}

\section{Introduction}
\label{sec:intro}

End-to-end (E2E) models~\cite{Ryan19,CC18,Graves12,RaoSakPrabhavalkar17,Chan15,chorowski2015attention,KimHoriWatanabe17,luscher2019rwth,miao2019online} 
have attracted large interest in both academia and industry. These models fold in components of the conventional automatic speech recognition (ASR) systems, namely an acoustic model (AM), pronunciation model (PM) and language model (LM), into a single neural network and optimize them jointly. E2E models simplify ASR system building and maintenance. They can have a much smaller model size than conventional ASR systems and are therefore more suitable for systems that perform the recognition on mobile devices. Among E2E variants, recurrent neural network transducer (RNN-T) \cite{Graves12} has shown potential for on-device streaming ASR \cite{Ryan19}.

Besides recognition quality, latency is another critical metric for streaming ASR. In this paper, we define recognition latency as the time difference between when the user stops speaking and when the system produces its final text hypothesis. It is desirable for model latency to be low enough that the system responds to the user quickly, while still high enough that it does not cut off the user's speech. Building models that have a better trade-off between word error rate (WER) and latency is crucial to achieving fast and accurate streaming speech recognition \cite{shuoyiin2019unified, chang2017endpoint, chang2018dcnn}

The decision of whether a user has stopped speaking is usually generated by an endpointer (EP) model. A voice activity detector (VAD) that detects speech and filters out non-speech is one such model. It can be used to declare an end-of-query (EOQ) as soon as VAD observes speech followed by a fixed interval of silence. VAD is not optimized to distinguish within-speech and query-end silences and may generate many false positive endpointing decisions. EOQ-based models address these issues \cite{MicCloser17}. They are directly optimized to distinguish speech and different types of silence including initial, intermediate and final. They have been shown to give better latency and WER trade-offs. 

Even with EOQ, the endpointer model and the ASR model are still optimized independently. Information captured by ASR models is not shared to the endpointer, which may be useful for making endpointing decisions. It would be better to optimize the endpointer and ASR models together. E2E models make this joint optimization simpler than with conventional modeling approaches. \cite{Shuoyiin19} does this by folding the EOQ detector into the RNN-T model by introducing a special token (\eos), signaling the end of speech, into RNN-T's output vocabulary. It is treated the same as all the other tokens during training. However, during inference it is used as one of the signals to end a search path. Premature \eos prediction may cause not only substitution errors but also deletions. 

To achieve better WER and latency trade-offs, we not only need the joint optimization of endpointer and ASR, the \eos token should also be predicted as close to the end of the last word as possible. In this work we propose to extend the joint RNN-T endpointer (EP) model \cite{Shuoyiin19} in a number of ways. First, we introduce penalties for emitting \eos too early or late in training, to encourage the model to find a good WER and latency trade-off. These penalties are applied to the \eos token, where the ground truth is obtained from a forced alignment between the transcript and audio signals. 

Second, premature \eos prediction causes a sequence level loss rather than a single token's. This leads us to explore whether sequence training \cite{kingsbury2009lattice,vesely2013sequence,rohit2018mwer} would address this problem. We hence investigated minimum word error rate \cite{rohit2018mwer} training for RNN-T EP models, which is found to yield both a WER and latency improvements. 
Third, we rescore RNN-T EP's hypotheses with a non-streaming model, namely Listen, Attend and Spell (LAS) \cite{sainath2019two}. The direct modeling of \eos in RNN-T makes the score combination with LAS, which emits \eos already, more consistent. While the rescoring model does not directly change the latency of RNN-T, WER gains it brings gives us more room for potential WER and latency trade-offs. The final setup, RNN-T EP with late penalty, LAS rescoring and MWER training, achieves a 18.7\% relative WER reduction and 40ms median latency and 160ms 90-percentile latency reductions on a Voice Search task comparing to the original RNN-T EP \cite{Shuoyiin19}.

The rest of the paper is organized as follows. Section \ref{sec:rnntep} explains the model architecture of the RNN-T EP and then details the proposed improvements of the RNN-T EP model using early and late penalties, MWER training and LAS rescoring. Section \ref{sec:exps} and \ref{sec:results} presents the experimental setup, results and analysis. 
\section{RNN Transducer and Endpointer}
\label{sec:rnntep}

\begin{figure}[t!]
\centering
  \includegraphics[width=\linewidth]{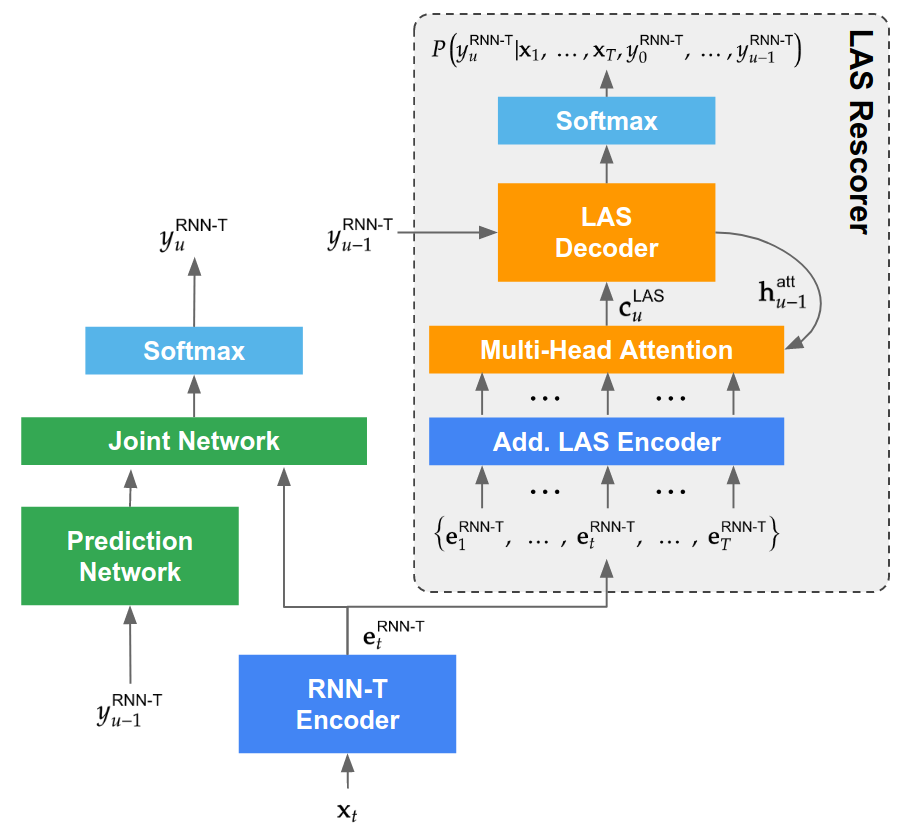} \\
  \caption{{Recurrent neural network transducer and endpointer (RNN-T EP) with non-streaming Listen, Attend and Spell (LAS) rescoring.}}
  \label{fig:rnntep}
\end{figure}

The recurrent neural network transducer and endpointer (RNN-T EP) model explored in this work is shown in Figure~\ref{fig:rnntep}. Let us denote the input acoustic frames as ${\bf x} = \{{\bf x}_1, \dots, {\bf x}_T\}$, where ${\bf x}_t \in \mathbb{R}^d$ are log-mel filterbank energies ($d=512$) and $T$ is the number of frames in ${\bf x}$. Each acoustic frame ${\bf x}_t$ is first passed through the RNN-T encoder, which consists of multiple layers of unidirectional LSTM layers. We denote the output of the RNN-T encoder as ${\bf e}_t$ and it is then forwarded to the RNN-T decoder for producing $y_t^\text{RNN-T}$. The output is decoded as soon as the input is encoded, without introducing additional latency incurred when processing the entire utterance at once. In this work, RNN-T is trained to directly predict word piece token sequence ${\bf y} = \{y_1, \dots, y_U\}$ where the last label $y_U$ is the special token \eos.

\subsection{Early and Late Penalties}

\begin{align}
\log {\mathbf P}_\text{RNN-T}(y_U | {\bf x}_t) \mathrel{{-}{=}} \big(& \max(0, ~\alpha_\text{early} * (t_{\eos} - t)) + \nonumber  \\ 
&\max(0, ~\alpha_\text{late} * (t - t_{\eos} - t_\text{buffer})) \big) \label{equ:earlylate}
\end{align}

Extending RNN-T's output vocabulary with a special token \eos helps improve its latency \cite{Shuoyiin19}, as the endpointing decision is made jointly with the model rather than with a separate endpointer. However, there is no constraint on when \eos should occur during training. A premature \eos prediction can result in deletion errors, while late predictions of \eos can increase latency as \eos is used to inform the system when the speech ends. In this paper, we address these issues by applying additional early and late penalties on the \eos token (Equation~(\ref{equ:earlylate})). Specifically, during training for every input frame in $\{{\bf x}_1, \dots, {\bf x}_T\}$ and every label $\{y_1, \dots, y_U\}$, RNN-T computes a $U \times T$ matrix ${\mathbf P}_\text{RNN-T}({\bf y}|{\bf x})$, which is used in the training loss computation. The last label $y_U$ is always \eos. We denote $t_{\eos}$ as the frame index after the last non-silence phoneme, obtained from the forced alignment of the audio with a conventional model. The RNN-T log-probability $\log {\mathbf P}_\text{RNN-T} (y_U|{\bf x})$ is modified to include a penalty at each time step $t$ for predicting \eos too early or too late. $t_\text{buffer}$ gives a grace period after the reference $t_{\eos}$ before the late penalty is applied. $\alpha_\text{early}$ and $\alpha_\text{late}$ are scales on the early and late penalties respectively. All hyper parameters are tuned experimentally.

\subsection{MWER Training}

Minimizing RNN-T loss corresponds to improving the log-likelihood of the training data. However, ASR system performance is measured in terms of WER, not log-likelihood. To address this mismatch, \cite{graves2014towards} proposes to minimize expected WER of the RNN-T model by approximating the expectation with samples draw from the model. Minimum word error rate training (MWER) is later applied to attention based LAS E2E models \cite{rohit2018mwer}. 

During the beam search decoding of the RNN-T EP model, the inference is terminated when either a blank symbol is generated at the last input frame or an \eos token is predicted. Premature \eos prediction results in deletion of the remaining reference target sequence, leading to a large sequence loss. This makes it more suitable for sequence training techniques. In this work, we hence investigate MWER training with N-best hypotheses for the RNN-T EP model.

\subsection{Listen, Attend and Spell Rescoring}
\label{sec:las_rescoring}

Non-streaming E2E models such as Listen, Attend and Spell (LAS) has shown better performance than streaming ones such as RNN-T. LAS has been explored to serve as a second pass rescorer\cite{sainath2019two}, that can still fit within the on-device latency constraints. As illustrated in Figure~\ref{fig:rnntep}, the model first collects the output of the RNN-T encoder of all the frames $\bf{e} = [\bf{e}_1, \dots, \bf{e}_T]$. They are then forwarded through an extra LAS encoder to generate a new set of encoder features for the LAS decoder. The decoder then computes output $\bf{y}_\text{LAS}$ accordingly. During inference, we first pick the top-K hypotheses from the RNN-T decoder. We then run the LAS model on each sequence in the teacher-forcing mode to compute a score, which combines log probability of the sequence and the attention coverage penalty \cite{chorowski2016towards}. The sequence with the highest LAS score is picked as the output sequence.

One of the issues in \cite{sainath2019two} is that RNN-T did not produce a score for \eos, while LAS is indeed trained to produce a score for it. Thus, when rescoring RNN-T hypotheses, an ``artificial" \eos score for RNN-T was added to the \eos from LAS. One can argue that including a score for \eos generated from RNN-T based on the inputs should help recognition, as it gives more confidence as to if the sentence should actually be completed. In this work, we look at improving LAS rescoring with the RNN-T EP model by including a score for \eos. The use of \eos token in RNN-T makes the score combination with LAS more consistent acorss all the output units. It is important to note that LAS rescoring cannot make the RNN-T model emit \eos faster; it can only improve the WER of RNN-T. However, the improvement of WER may provide additional room to trade WER for latency.

\section{Experimental Setups}
\label{sec:exps}

\subsection{Dataset}

We use the same multidomain dataset as \cite{narayanan2018toward} for training. Multistyle training (MTR) is used for noise robustness \cite{kim2017generation}. During training, a noise configuration, which defines mixing conditions like the size of the room, reverberation time, position of the microphone, speech and noise sources, signal to noise ratio (SNR), etc, for each utterance is randomly sampled from a collection of 3 million pre-generated configurations. The detailed noise configuration can be found in \cite{narayanan2018toward}. 
The test set we use consists of 14K Voice Search utterances with duration less than 5.5 seconds long. They are all anonymized and hand-transcribed, and are representative of Google traffic.

\subsection{Modeling}

The input waveforms are framed using a 32 msec window with 10 msec shift. Globally normalized 128 dimension logmel features extracted from frequencyies spanning from 125 Hz to 7.5kHz are used as inputs. The input window size is 4, consisting of 3 frames on the left and no future context. It is further subsampled by a factor of 3 making the system operate at 33 Hz \cite{pundak2016lower}.

Similar to \cite{li2018multi}, multidomain models are trained with domain id as an additional input for learning domain-dependent variations. Following \cite{Ryan19}, all LSTM layers in the model are unidirectional, with 2048 units and a projection layer with 640 units. The RNN-T encoder consists of 8 LSTM layers, with a time-reduction layer after the second layer. The RNN-T decoder consists of a prediction network with 2 LSTM layers, and a joint network with a single feed-forward layer with 640 units. The additional LAS encoder consists of 2 LSTM layers. The LAS decoder consists of multi-head attention \cite{vaswani2017attention} with 4 attention heads, which is fed into 2 LSTM layers. All models are trained on 8x8 Cloud TPU using the Tensorflow Lingvo toolkit \cite{shen2019lingvo} to predict 4,096 word pieces including the \eos token. 

\subsection{Inference}
Despite the use of multidomain training, this work focuses only on the Voice Search task. We append the \eos token only to the Voice Search queries and keep the other data untouched. We report both the recognition performance in terms of word error rate (WER) and the latency of the models for Voice Search only. The latency metrics used in this paper includes median latency (EP50), 90 percentile latency (EP90) and the endpointing coverage (EOU) which represents the percentage of the test data actually receives an end-of-utterance signal from the endpointer model.

There is a trade-off between accuracy and latency, which is often depicted by ROC curves. For EOU EPs, it is obtained by adjusting the endpointing decision threshold. For RNN-T EPs, the endpointing decision is defined by:
\begin{equation}
p(\eos | {\bf x}_1, \dots, {\bf x}_t, y_0^\text{RNN-T}, \dots, y_{t-1}^\text{RNN-T})^{\alpha_{\eos}} \ge \beta \label{equ:eos_penalty}.
\end{equation}
$\alpha_{\eos}$ is a penalty term for the posterior of \eos that modifies the ordering for the hypothesis with \eos. $\beta$ is a predefined threshold that determines if \eos is allowed in the search beam \cite{Shuoyiin19}. Sweeping $\alpha_{\eos}$ and $\beta$ gives us a ROC curve of the WER and latency trade-off. For simplicity, we most of the time report a single trade-off point and only show the ROC curves at the end for the final comparisons. 
\section{Results}
\label{sec:results}

\subsection{Baseline}

We first train a RNN-T model to predict 4,096 word pieces for the ASR task only (no \eos) as was done in past \cite{Ryan19}. This RNN-T cannot be used to output an endpointing decision and an external EOQ EP is used \cite{MicCloser17, Shuoyiin19} (B1 in Table~\ref{tbl:baseline}). The endpointer and the RNN-T ASR model are trained independently and at the inference time, the information from RNN-T's hypotheses cannot be used for endpointing decisions. To address this issue, we also trained a joint endpointing and recognition RNN-T EP model proposed in \cite{Shuoyiin19} (B2 in Table~\ref{tbl:baseline}). As suggested in \cite{Shuoyiin19}, we also use the independnetly trained EOQ EP as a backup for the RNN-T EP model. From Table~\ref{tbl:baseline}, the RNN-T EP (B2) shows good latency gains (130ms EP50 and 200ms EP90 latency reductions and a 5.4\% absolute EOU coverage improvement) but has an increase of 0.3\% WER. One assumption of this regression is that during training \eos is treated the same as all the other tokens, with no constraint on how early or late \eos should occur; however in inference, a path ends when a \eos token is predicted. Predicting EOS prematurely brings in deletion errors.

\begin{table}[t!]
\caption{Quality and latency performance of the baseline models.}
\centering
\begin{tabular}{l|c|c|c|c}
\toprule
\multirow{2}{*}{\tbh{Exp.}} & \tbh{WER} & \tbh{EP50} & \tbh{EP90} & \tbh{EOU}  \\
~ & \scriptsize{(\%)} & \scriptsize{(ms)} & \scriptsize{(ms)} & \scriptsize{(\%)} \\
\midrule
\midrule
\tbh{B1} \scriptsize{RNN-T} & {\bf 7.2} & 540 & 910 & 86.7 \\
\midrule
\tbh{B2} \scriptsize{RNN-T EP} & 7.5 & {\bf 410} & {\bf 710} & {\bf 92.1} \\
\bottomrule
\end{tabular}
\label{tbl:baseline}
\end{table}

\subsection{Early and Late Penalties}

To address the potential premature \eos prediction, we adopt an early penalty term to the training. It is added only if \eos is predicted at any frame earlier than its ground truth time. When adding the early penalty, we scale it by a factor of 0.1 which is found to work well. This (E1 in Table~\ref{tbl:earlylate}) reduces the WER from 7.5\% to 7.2\% but degrades on latency comparing to B2. 
The use of early penalty does help the model to address premature \eos prediction but has the risk of the model learning to over-delay its predictions, which leads to worse latency. The regression on EP90 is more sever which is because many tail cases are not endpointed by RNN-T EP and they simply fall back to the EOQ EP. 

We further introduce a late penalty term to penalize the \eos prediction that happens too late comparing to the ground truth. 
During training the granularity of the time is frame (particularly 60ms in our setup). We experimented with $t_\text{buffer} = \{3, 5, 7\}$ which corresponds to a grace period of 180ms, 300ms and 420ms after the reference \eos label. The results are presented in Table~\ref{tbl:earlylate}. With 3 frames' buffer, we obtain the best median latency but 5-frame gives the best 90-percentile latency which is still worse than B2. We take model E2, namely the RNN-T EP model with early penalty and 3-frame late penalty, as the setup for following experiments.

\begin{table}[t!]
\caption{Quality and latency performance of models with early and late penalties.}
\centering
\begin{tabular}{l|c|c|c|c}
\toprule
\multirow{2}{*}{\tbh{Exp.}} & \tbh{WER} & \tbh{EP50} & \tbh{EP90} & \tbh{EOU}  \\
~ & \scriptsize{(\%)} & \scriptsize{(ms)} & \scriptsize{(ms)} & \scriptsize{(\%)} \\
\midrule
\midrule
\tbh{E1} \scriptsize{Early} & 7.2 & 430 & 830 & 90.7 \\
\midrule
\tbh{E2} \scriptsize{E1 + 3Frame\_Late} & 7.2 & {\bf 380} & 850 & 88.3 \\
\tbh{E3} \scriptsize{E1 + 5Frame\_Late} & 7.2 & 400 & {\bf 790} & {\bf 91.5} \\
\tbh{E4} \scriptsize{E1 + 7Frame\_Late} & 7.2 & 540 & 860 & 90.8 \\
\bottomrule
\end{tabular}
\label{tbl:earlylate}
\end{table}

\subsection{MWER training}

For the RNN-T EP model, a wrong prediction of \eos leads to not just a token error but a sequence level loss as it is used to terminate a path in beam search. MWER training optimizes sequence level loss and penalizes WER when \eos is emitted too early, thus prompting our investigation in this section. 

We conducted MWER training for the RNN-T model without \eos (B1) and the best RNN-T EP (E2). Both the pre- and post-MWER results are reported in Table~\ref{tbl:embr}. For B1, the latency is controlled by a separate EOU EP and hence remains the same after MWER training (B3). But the WER reduces from 7.2\% to 6.9\%. While for E2, MWER training (E4) maintains the same 7.2\% WER as E2 but achieves 220ms EP90 reduction with 50ms regression on EP50. 
Because optimizing MWER already penalizes premature \eos predictions, we turn off the early penalty for MWER training. This (E5 in Table~\ref{tbl:embr}) reduces WER from 7.2\% to 6.9\% WER and more importantly it still yields a 270ms EP90 latency reduction while maintaining the same EP50 latency as E2. MWER training of the RNN-T EP model with only late penalty can bring in both WER and latency improvements. Comparing to B2, E5 gives 8.0\% relative WER reduction and 30ms EP50 and 130ms EP90 latency reductions.

\begin{table}[t!]
\caption{Quality and latency performance of models w and w/o MWER training.}
\centering
\begin{tabular}{l|c|c|c|c}
\toprule
\multirow{2}{*}{\tbh{Exp.}} & \tbh{WER} & \tbh{EP50} & \tbh{EP90} & \tbh{EOU}  \\
~ & \scriptsize{(\%)} & \scriptsize{(ms)} & \scriptsize{(ms)} & \scriptsize{(\%)} \\
\midrule
\midrule
\tbh{B1} \scriptsize{RNN-T} & 7.2 & 540 & 910 & 86.7 \\
\tbh{B2} \scriptsize{RNN-T EP} & 7.5 & 410 & ~710 & 92.1 \\
\midrule
\tbh{B3} \scriptsize{B1 + MWER} & {\bf 6.9} & 540 & 910 & 86.7 \\
\midrule
\midrule
\tbh{E2} \scriptsize{B2 + Early + Late} & 7.2 & {\bf 380} & 850 & 88.3 \\
\midrule
\tbh{E4} \scriptsize{E2 + MWER} & 7.2 & 430 & 630 & {\bf 97.3} \\
\tbh{E5} \scriptsize{E4 - Early} & {\bf 6.9} & {\bf 380} & {\bf 580} & 95.5 \\
\bottomrule
\end{tabular}
\label{tbl:embr}
\end{table}

\begin{table}[t!]
\caption{Quality and latency performance of models with 2nd pass LAS rescoring.}
\centering
\begin{tabular}{l|c|c|c|c}
\toprule
\multirow{2}{*}{\tbh{Exp.}} & \tbh{WER} & \tbh{EP50} & \tbh{EP90} & \tbh{EOU}  \\
~ & \scriptsize{(\%)} & \scriptsize{(ms)} & \scriptsize{(ms)} & \scriptsize{(\%)} \\
\midrule
\midrule
\tbh{B2} \scriptsize{RNN-T EP} & 7.5 & 410 & ~710 & 92.1 \\
\tbh{E2} \scriptsize{B2 + Early + Late} & 7.2 & 380 & 850 & 88.3\\
\midrule
\tbh{E6} \scriptsize{E2 + LAS} & 6.4 & 380 & 850 & 88.3 \\
\quad~ \scriptsize{+ re-sweep} & 6.4 & 370 & 740 & 91.4 \\
\quad~ \scriptsize{+ ignore RNN-T \eos score} & 6.6 & 370 & 740 & 91.4 \\
\midrule
\tbh{E7} \scriptsize{E6 + MWER LAS only} & 6.2 & {\bf 350} & 620 & 92.4 \\
\tbh{E8} \scriptsize{E7 + MWER All} & {\bf 6.1} & 370 & {\bf 550} & {\bf 95.2} \\
\bottomrule
\end{tabular}
\label{tbl:las}
\end{table}

\subsection{LAS Rescoring}

So far we see good latency reductions, but the WER gains are small. In the literature, two-pass model that runs RNN-T as the first pass streaming model for fast response and LAS as the rescorer has been shown to be effective in WER reductions. We hence investigate the effect of LAS rescoring on RNN-T EP model. We took the pre-MWER model E2 and added an additional encoder with two LSTM layers and an extra LAS decoder (Figure~\ref{fig:rnntep}). 
They are trained with cross entropy (CE) loss with the RNN-T weights frozen. The results are presented as E6 in Table~\ref{tbl:las}. In this work, the latency is only measured for the first pass model. With the same decoding configuration as E2, LAS rescoring reduces the WER by 11.1\% relative from 7.2\% to 6.4\%. 
Although LAS rescoring cannot directly affect first pass latency, with the WER gains, we may be able to trade WER for latency. We further swept the penalty scale for \eos and obtained an operation point with the same 6.4\% WER but 10ms EP50 and 130ms EP90 reductions. As mentioned in Section~\ref{sec:las_rescoring}, one problem for LAS rescoring of RNN-T without \eos as done in \cite{sainath2019two} is that RNN-T does not generate an explicit \eos score to combine with that from LAS. To simulate that effect, we zeroed out the \eos score from RNN-T EP and swept a global value to be combined with LAS \eos score. The result (E6 + ignore RNN-T \eos score in Table~\ref{tbl:las}) shows an increase in WER from 6.4\% to 6.6\%, highlighting the benefit of using \eos in RNN-T EP for LAS rescoring. 

Instead of CE loss, MWER loss of RNN-T outputs can be used to update the LAS rescorer (E7 in Table~\ref{tbl:las}). It further reduces the WER down to 6.2\% and obtains 100ms EP90 reductions. Moreover, when we update both RNN-T and LAS during MWER (E8), we can obtain another 70ms EP90 reduction. Comparing to B2, the RNN-T EP + LAS with MWER training gives a 18.7\% relative WER reduction and 40ms EP50 and 160ms EP90 reductions.

\subsection{Analysis}

\begin{figure}[t!]
\hspace{-0.025\linewidth}
\includegraphics[width=1.05\linewidth]{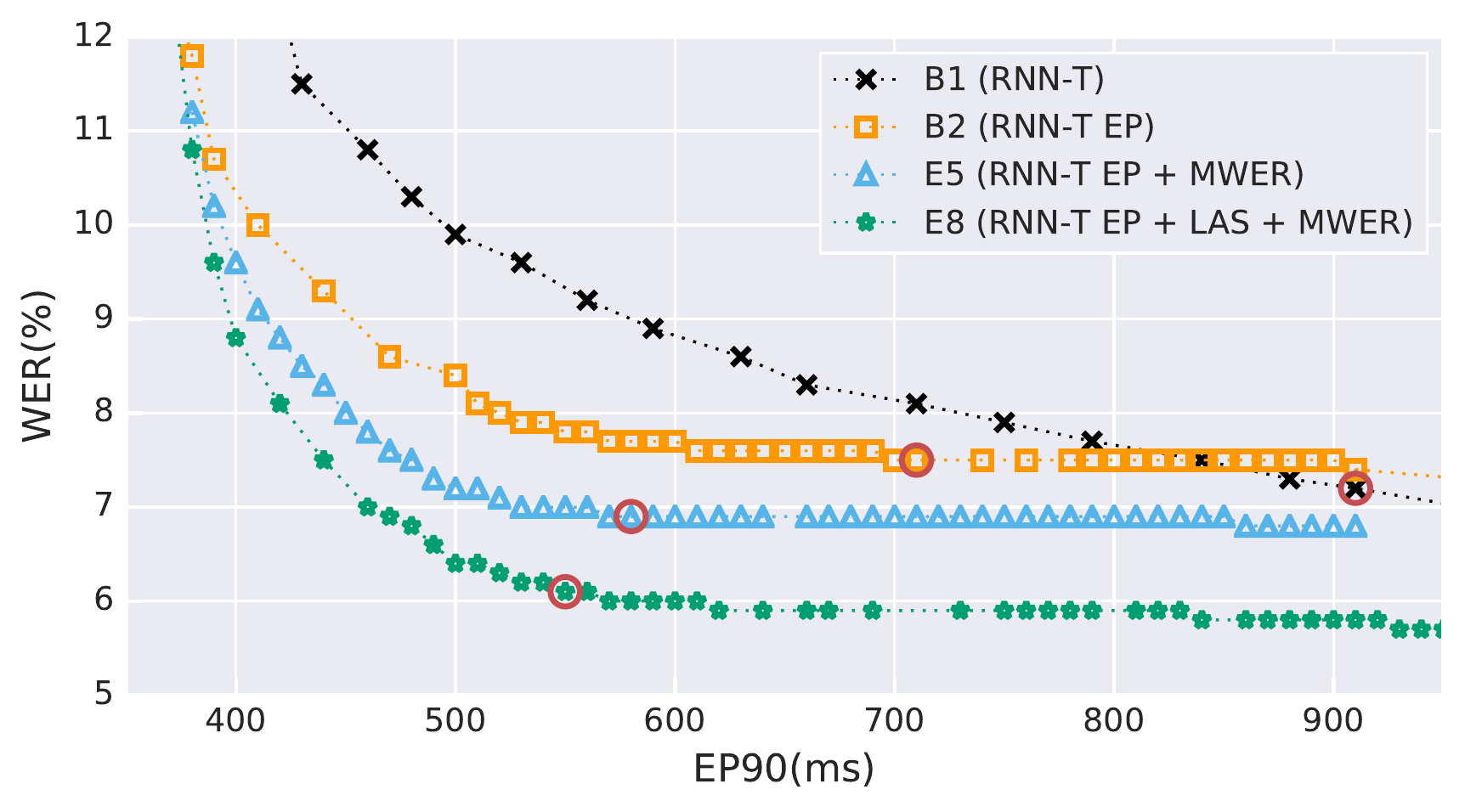} \\
\caption{{ROC curves of WER and 90-percentile latency (EP90) trade-offs for RNN-T (B1), the original RNN-T EP (B2), the proposed RNN-T EP with late penaly and MWER training (E5) and with additional LAS rescoring (E8). Red circles represent operation points reported in early sections.}}
\label{fig:roc_ep90}
\vspace{-0.1in}
\end{figure}

The proposed RNN-T EP with late penalty and MWER training (E5) gives us both WER and latency improvement over the RNN-T (B1) and the original RNN-T EP (B2). Further WER improvement is achieved via a second pass LAS rescorer (E8). In this section, we compare these systems across different operating points. We plotted the WER vs latency (EP90) curve for these four models (B1, B2, E5, E8) in Figure~\ref{fig:roc_ep90} by varying the penalty scale $\alpha_\eos$ and threshold $\beta$. Lower curves are better. RNN-T (B1) tends to delay outputs and has worse latency. With \eos, RNN-T EP (B2) addresses the latency problem but with some WER degradations. With the modifications proposed in this work, namely late penalty, MWER and LAS rescoring, both E5 and E8 have much better WER and latency trade-offs.

\vfill\pagebreak

\bibliographystyle{IEEEbib}
\bibliography{refs}

\end{document}